\documentclass[useAMS,usenatbib,usegraphicx]{mn2e}

\title[Modelling the Warm H$_2$ Emission of Cometary Knots] {Modelling the Warm H$_2$ Infrared Emission of the Helix Nebula Cometary Knots}

\author[I. Aleman et al.]
{Isabel Aleman$^{1,2}$\thanks{E-mail: isabel@astro.iag.usp.br},
Albert A. Zijlstra$^{1}$, 
Mikako Matsuura$^{3,4}$, 
Ruth Gruenwald$^{2}$, 
\newauthor % starts a new line in the author environment
and Rafael K. Kimura$^{2}$\\
$^{1}$Jodrell Bank Centre for Astrophysics, The Alan Turing Building, 
      School of Physics and Astronomy, The University of Manchester, \\ 
      Oxford Rd, Manchester, M13 9PL, UK\\
$^{2}$IAG-USP, Universidade de S\~ao Paulo, Cidade Universit\'aria, 
      Rua do Mat\~ao 1226, S\~ao Paulo, SP, 05508-090, Brazil \\
$^{3}$Institute of Origins, Astrophysics Group, Department of 
      Physics and Astronomy, University College London, Gower Street, \\
      London, WC1E 6BT, UK\\
$^{4}$Institute of Origins, Mullard Space Science Laboratory, 
      University College London, Holmbury St. Mary, Dorking, Surrey,\\ 
      RH5 6NT, UK}

\begin{document}

\date{Accepted 2011 May 17.  Received 2011 May 16; in original form 2010 September 28 
}

\pagerange{\pageref{firstpage}--\pageref{lastpage}} \pubyear{2011}

\maketitle

\label{firstpage}

%========================================================================
\begin{abstract}
%========================================================================

Molecular hydrogen emission is commonly observed in planetary nebulae. Images taken in infrared H$_2$ emission lines show that at least part of the molecular emission is produced inside the ionised region. In the best-studied case, the Helix nebula, the H$_2$ emission is produced inside cometary knots (CKs), comet-shaped structures believed to be clumps of dense neutral gas embedded within the ionised gas. Most of the H$_2$ emission of the CKs seems to be produced in a thin layer between the ionised diffuse gas and the neutral material of the knot, in a mini photodissociation region (PDR). However, PDR models published so far cannot fully explain all the characteristics of the H$_2$ emission of the CKs. In this work, we use the photoionisation code \textsc{Aangaba} to study the H$_2$ emission of the CKs, particularly that produced in the interface H$^+$/H$^0$ of the knot, where a significant fraction of the H$_2$ 1-0~S(1) emission seems to be produced. Our results show that the production of molecular hydrogen in such a region may explain several characteristics of the observed emission, particularly the high excitation temperature of the H$_2$ infrared lines. We find that the temperature derived from H$_2$ observations even of a single knot, will depend very strongly on the observed transitions, with much higher temperatures derived from excited levels. We also proposed that the separation between the H$\alpha$ and [\mbox{N\,{\sc ii}}] peak emission observed in the images of CKs may be an effect of the distance of the knot from the star, since for knots farther from the central star the [\mbox{N\,{\sc ii}}] line is produced closer to the border of the CK than H$\alpha$.

\end{abstract}


%========================================================================
\section{Introduction}
%========================================================================

Planetary nebulae (PNe) are formed by the ejection of the outer layers of low and intermediate mass stars (1 to 8 $M_{\sun}$) in their final stages of evolution. The ejecta originate while the star is on the Asymptotic Giant Branch. Afterwards, the rapidly heating star drives a molecular dissociation front and an ionisation front, which eventually overrun the ejecta. The study of the composition and structure of PNe, in particular, the molecular gas (its location, distribution, and physical conditions), can provide information about the earlier evolutionary stages \citep{Davis_etal_2003, Volk_etal_2004}, as well as on the physics of dissociation and ionisation fronts \citep{Henney_etal_2007}.

Infrared (IR) emission of H$_2$ has been identified in more than 70 PNe \citep{Hora_etal_1999, Sterling_etal_2008}, primarily bipolar nebulae \citep{Kastner_etal_1996}. The precise location of the observed H$_2$ emission and the dominant excitation mechanism are still under debate \citep{Hora_etal_1999, Likkel_etal_2006, Henney_etal_2007}.

Analysis of observations and models of the H$_2$ IR line emission of PNe indicate that the emission is produced both within the neutral envelope and within the ionised region \citep[][and references therein]{Aleman_Gruenwald_2004}. \citet{Aleman_Gruenwald_2011} showed that, in cases of high temperature central stars, a significant part of the H$_2$ emission can be produced by the diffuse gas in the H$^+$/H$^0$ transition region of PNe.

The Helix nebula (NGC 7293) is an evolved PN excited by a T$_{\star}$ = 120000~K, L$_{\star}$ = 100 L$\sun$ central star \citep{Henry_etal_1999}. High-resolution images of this PN have shown that the H$_2$ emission arises from its large population of globules, the so-called cometary knots (CKs), embedded in the ionised gas \citep[e.g.][]{Matsuura_etal_2009}. Many authors argue that CKs are largely responsible for the H$_2$ emission produced within the ionised region \citep{Beckwith_etal_1978, Gussie_Pritchet_1988, Reay_etal_1988, Tielens_1993, Schild_1995, Speck_etal_2002, Matsuura_etal_2009}.

CKs are structures that resemble comets, particularly in images taken in H$\alpha$, [\mbox{N\,{\sc ii}}] $\lambda$6583, and H$_2$ 1-0~S(1) lines. Knots and filamentary condensations are also seen in other PNe \citep{ODell_etal_2002} in the transition region between the ionised and neutral H. In the Helix nebula, the bright cusp points towards the central star and the tail in the opposite direction, which can indicate that the excitation is connected with the central star. Analyses of the emission of these structures suggest that they are significantly denser than the gas around them. While the typical diffuse gas density is around 10$^2$-10$^4$ cm$^{-3}$ \citep{Osterbrock_Ferland_2006, Kwok_2000}, in the CKs the density is around 10$^5$-10$^6$ cm$^{-3}$ \citep{Huggins_etal_1992, Meaburn_etal_1998, Matsuura_etal_2007}. The origin of the CKs is still uncertain. They may be formed within the original wind of the progenitor star or by instabilities and fragmentation of the swept-up shell during the onset of the PN phase. \citet{Huggins_Frank_2006} show that the latter is better supported by the observed properties of the knots. \citet{Matsuura_etal_2009} suggest a possible origin within a spiral density wave in the wind caused by an orbiting companion.

There is no evidence that the H$_2$ emission in CKs is produced by shocks \citep{Huggins_etal_2002, Odell_etal_2005, Matsuura_etal_2007, Matsuura_etal_2008}. On the other hand, as mentioned above, the emission of the CKs seems to be linked to the central star radiation field \citep{Odell_etal_2005, Odell_etal_2007}. The H$_2$ emission is more intense in a thin layer in the surface of the CKs towards the central star. However, analysis based on traditional models of photodissociation regions (PDRs) were unable to reproduce the high excitation temperature of the H$_2$ emission ($\sim$900~K) estimated from the observations \citep{Huggins_etal_2002, Odell_etal_2007}. Recently, \citet{Henney_etal_2007} showed that advection can cause the ionisation and dissociation front to merge, leading to enhanced heating of the molecular gas and reproducing well the excitation temperatures of H$_2$. It is evident from the work of Henney et al. that the physical conditions in the interface between the ionized diffuse gas and the CK are of key importance for the H$_2$ IR emission. The high excitation temperature of the H$_2$ emission could be naturally explained if part of the H$_2$ emission is produced in the H$^+$/H$^0$ transition of the CK.

In the present paper, the ionised and partially ionised regions of CKs are modelled with the one-dimensional photoionisation code \textsc{Aangaba}. Our aim is to study in more detail the H$_2$ emission in the H$^+$/H$^0$ interface of the CK. A grid of models is obtained to study how the H$_2$ IR emission depends on different interface density profiles, CK radius, distance from the ionizing source, and dust-to-gas ratio. We compare the H$_2$ IR emission with the atomic emission. Our calculations also provide good estimates of the H$_2$ formation and destruction rates inside the CKs, which indicate how long the molecule may survive inside the CKs and whether it forms in-situ or may have survived the earlier evolutionary phase.

We use our models to study the CKs in the Helix nebula (NGC 7293). The Helix is one of the nearest PNe (219 pc), and high resolution images resolved the structure of the CKs \citep{Odell_etal_2004, Meixner_etal_2005, Hora_etal_2006, Matsuura_etal_2007, Matsuura_etal_2008, Matsuura_etal_2009}. The detailed observations allow us to test the model. Our models are described in Section 2. The results are discussed in Section 3 and conclusions are summarized in Section 4.

%========================================================================
\section{Models}
%========================================================================

In Section 2.1, a brief description of the photoionisation code Aangaba is given. The parameters assumed for the Helix nebula model are discussed in Section 2.2. A description of how the CKs were simulated is given in Section 2.3.

%************************************************************************
\subsection{General Description of the Code}
%************************************************************************

The \textsc{Aangaba} code simulates the physical conditions inside a photoionised nebula, given the ionising spectrum, as well as the characteristics and distribution of gas and dust \citep{Gruenwald_Viegas_1992}. Since the gas and dust temperatures and densities depend on each other, as well as on the position inside the nebula, the code makes iterative calculations across the nebula. The code begins the calculations at the inner border of the nebula and continues the calculation in steps in the outward direction. In each location, the physical conditions, as for example the ionic and molecular density, the electronic density and temperature, the dust temperature, and continuum and line emission, are calculated. The calculation ends when the code reaches the limit chosen by the user (for example, a given ionisation degree, gas temperature, optical depth, etc.). Geometrical dilution and extinction of the radiation by gas and dust are taken into account. The transfer of the primary and diffuse radiation fields is treated in the outward-only approximation\textbf{\footnote{\textbf{The outward-only is an usual approximation to treat the diffuse radiation in one-dimensional photoionization codes \citep[see, for example,][]{Pequignot_etal_2001, Stasinska_2009}. It was introduced by \citet{Tarter_Salpeter_1969}. In this approximation, in the words of \citet{Pequignot_etal_2001}, `all the diffuse photons are assumed to be emitted isotropically in the outer half-space'. The diffuse radiation photons emitted in the inward direction are assumed to travel to the point symmetrically opposite with respect to the central star in the other side of the nebula through the optically thin gas without being absorbed. This allows the code to proceed with the calculation in the outward direction only, that is, it does not need to recalculate previous steps to account for the diffuse radiation emitted more externally.}}}.

Twelve elements and their ions are included in the code: H, He, C, N, O, Mg, Ne, Si, S, Ar, Cl, and Fe. The species H$_2$, H$_2^+$, H$_3^+$, and H$^-$ are also included in the code \citep[H$_3$ is not included because it is unstable;][]{Bordas_etal_1990}. The density of each ion of each species is calculated with the assumption of chemical and ionisation equilibrium. The processes of ionization, recombination and charge exchange are taken into account in the equilibrium equations of the atomic species. For the H bearing species, forty reactions of formation and destruction are included in the chemical equilibrium equations \citep{Aleman_Gruenwald_2004}.

The population of the H$_2$ rovibrational levels of the three lowest electronic bound states is calculated by assuming statistical equilibrium, i.e., the total population rate of a level is equal to its total depopulation rate. For the electronic ground level, several excitation and de-excitation mechanisms (radiative and collisional), as well as the possibility that H$_2$ is produced or destroyed in any given level, are included. For upper electronic levels, only radiative electronic transitions between each upper state and the ground state are included, since this must be the dominant mechanism. The population of the H$_2$ rovibrational levels of the electronic ground state by radiative mechanisms occurs through two main routes: electric quadrupole transitions between the rovibrational levels, involving IR photons, or electric dipole transitions to upper electronic states with subsequent decay to the ground state, involving UV photons (UV pumping). Collisions may also change the H$_2$ energy level. We included collisions of H$_2$ with the main components of the gas, i.e., H, H$^+$, He, H$_2$, and electrons. In this work we are interested in the lines produced by the rovibrational transitions of H$_2$, whose wavelengths are in the 0.28~$\mu$m to 6.2~mm range of the electromagnetic spectrum. More details about the calculation of the H$_2$ level population can be found in \citet{Aleman_Gruenwald_2011}.

The gas temperature is calculated assuming thermal equilibrium, that is, the total input of energy in the gas per unit time and volume is balanced by the total loss of energy per unit time and volume. Several mechanisms of gain and loss of energy by the gas due to atomic species, dust, and H$_2$ are taken into account. The gas heating mechanisms are photoionization of atoms, atomic ions, and H$_2$ by the primary and diffuse radiation; H$_2$ photodissociation (direct and two steps); H$_2$ formation on grain surfaces, by associative detachment, and by charge exchange with H; H$_2$ collisional de-excitation; and photoelectric effect on dust surfaces, while the cooling mechanisms are emission of collisionaly excited lines; radiative and dielectronic atomic recombination; thermal collisional atomic ionization; free-free emission; collisional excitation of H$_2$; destruction of H$_2$ by charge exchange with H$^+$; H$_2$ collisional dissociation; and collision of gas-phase particles with dust grains.\textbf{Models by \citet{Tielens_Hollenbach_1985} and \citet{Spaans_Meijerink_2005} indicate other molecules become important contributors to the thermal balance if $A_v >$ 3.5. Since in our models $A_v <$ 2.7, we do not include molecules other than H$_2$ in the thermal equilibrium.}

%************************************************************************
\subsection{The Helix Model}
%************************************************************************

The parameters for the Helix nebula model are given in Table 1. We assume that the central star radiates as a blackbody with $T_{\star}=$ 120000~K and $L_{\star}=$ 100 $L_{\sun}$ \citep{Henry_etal_1999, Odell_etal_2007}. The density of the diffuse gas is $n_D =$ 50 cm$^{-3}$ \citep{Meixner_etal_2005}.

Elemental abundances were determined by \citet{Henry_etal_1999} for He, O, C, N, Ne, S, and Ar, from Helix line emission observations and ICFs obtained from photonisation models. The C/O ratio obtained by \citet{Henry_etal_1999} is 0.87. The detection of C-bearing molecules, such as H$_2$CO, c-C$_3$H$_2$, and C$_2$H, has been used to argue that the Helix may be in fact C-rich \citep{Tenenbaum_etal_2009}. We ran some models with C-rich abundances, but we found it does not affect our results for H$_2$ significantly, and will assume the \citet{Henry_etal_1999} abundances. For the remaining elements taken into account in the code, Mg, Si, Cl, and Fe, we adopt the values from \citet{Stasinska_Tylenda_1986}.

\citet{Speck_etal_2002} model the IR emission of Helix and found the dust-to-gas ratio $M_\mathrm{d}/M_\mathrm{g}=$ 10$^{-3}$, which is an average value for PNe \citep{Stasinska_Szczerba_1999}. Our models assume by default amorphous carbon grains with 0.1$\mu$m radius. We also did calculations with silicate dust, but we found \citep[as also discussed in ][]{Aleman_Gruenwald_2004} that the choice of compound does not cause significant changes in the H$_2$ density.

We assume a distance of 219 pc \citep{Harris_etal_2007, Benedict_etal_2009}, obtained from measurements of the central star parallax.

%-----------------------------------------------------
\begin{table}
\centering
\begin{minipage}{140mm}
\caption{Input Parameters for the Helix Model.} 
\begin{tabular}{@{}ll@{}}
\hline
Parameter & Value\footnote{References are given in the text.}\\
\hline
$T_{\star}$  & 120000~K         \\
$L_{\star}$  & 100 $L_{\sun}$  \\
$n_{D}$      & 50 cm$^{-3}$     \\
Dust material& amorphous carbon \\
Grain radius & 0.1 $\mu$m       \\
Distance     & 219 pc           \\
\hline
Element      & Abundances (relative to H, by number) \\
\hline
He           & 1.20$\times10^{-1}$   \\
O            & 4.60$\times10^{-4}$   \\
C            & 4.00$\times10^{-4}$   \\
N            & 2.48$\times10^{-4}$   \\
Ne           & 1.52$\times10^{-4}$   \\
S            & 1.48$\times10^{-6}$   \\
Ar           & 3.10$\times10^{-6}$   \\
Mg           & 3.80$\times10^{-7}$   \\
Si           & 3.50$\times10^{-7}$   \\
Cl           & 3.20$\times10^{-9}$   \\
Fe           & 4.70$\times10^{-7}$   \\
\hline
\label{TabHelixPar}
\end{tabular}
\end{minipage}
\end{table}
%-----------------------------------------------------

%************************************************************************
\subsection{Simulating the Cometary Knots}
%************************************************************************

The CKs are simulated as an increase in the density profile of the Helix nebula model at a given distance. We construct a grid of CK models with different core densities ($n_K$), density profiles, dust-to-gas ratios ($M_d/M_g$), and distances from the central star ($R_K$). We obtain models with $n_K$ between 10$^5$ to 10$^6$ cm$^{-3}$ \citep{Huggins_etal_1992, Meaburn_etal_1998, Matsuura_etal_2007}. We assume that the density profile has an increase from $n_D$ to $n_K$ over a given distance $\Delta R$. In the following discussions, we call this region the interface of the knot. The region where the density reached the maximum value ($n_K$) is called the core. We study four types of density profiles: step function (hereafter type 1), linear (type 2), $r^2$ (type 3), and exponential (type 4). According to \citet{Odell_Handron_1996}, the Helix CKs have $M_d/M_g$ between 7$\times10^{-4}$ and 7$\times10^{-2}$. We obtain models for $M_d/M_g$ of 10$^{-3}$ and 10$^{-2}$. In each model, the dust-to-gas ratio and the chemical composition are assumed the same for CKs and diffuse gas. The parameters of the CK models discussed in the present work are listed in Table \ref{TabKnotPar}.

%-----------------------------------------------------
\begin{table}
\centering
\begin{minipage}{140mm}
\caption{Parameters of the CKs models.}
\begin{tabular}{@{}lccccc@{}}
\hline
Model & $R_K$ (arcsec) & $\Delta R$ (arcsec) & $n_K$ (cm$^{-3}$) &  
$M_d/M_g$ & Profile\\
\hline
K1  & 129 & 0.0  & 10$^5$ & 10$^{-3}$ & 1\\
K2  & 129 & 0.2  & 10$^5$ & 10$^{-3}$ & 2\\
K3  & 129 & 0.2  & 10$^5$ & 10$^{-3}$ & 3\\
K4  & 129 & 0.2  & 10$^5$ & 10$^{-3}$ & 4\\
K5  & 129 & 0.01 & 10$^5$ & 10$^{-3}$ & 4\\
K6  & 129 & 0.5  & 10$^5$ & 10$^{-3}$ & 4\\
K7  & 129 & 0.2  & 10$^5$ & 10$^{-2}$ & 4\\
K8  & 129 & 0.2  & 10$^6$ & 10$^{-3}$ & 4\\
K9  & 210 & 0.2  & 10$^5$ & 10$^{-3}$ & 4\\
K10 & 383 & 0.2  & 10$^5$ & 10$^{-3}$ & 4\\
K11 & 450 & 0.2  & 10$^5$ & 10$^{-3}$ & 4\\
K12 & 501 & 0.2  & 10$^5$ & 10$^{-3}$ & 4\\
\hline
\label{TabKnotPar}
\end{tabular}
\end{minipage}
\end{table}
%-----------------------------------------------------

The code starts the calculations in the inner border of the nebula and continues outward along the radial direction. At a given distance ($R_K$), an increase in density simulates the CK. For the present work, we calculate models with $R_K =$ 130 to 500 arcsec (see discussion in Sect. \ref{ResHelixMod}). We stop the calculations where the gas temperature, which decreases with distance from the central star, reaches 100~K. We define $\Delta R_T$ as the distance from the border of the knot to this point.

An IDL routine was developed to simulate a three-dimensional CK, allowing the calculation of surface brightness of atomic and H$_2$ lines by the integration of the emissivity along the line of sight inside the CK. In this routine, the knot is assumed to have cylindrical symmetry, with a semi-spherical head pointing in the direction of the central star. The CK is assumed to be seen edge-on. We assume that the one-dimensional profile of the emissivity, which is calculated with \textsc{Aangaba}, is the same for every line along the direction parallel to the symmetry axis, starting from the border of the knot. This is a simple approximation to simulate a comet-shaped knot, but it allows us to obtain a reasonable estimate of the H$_2$ lines surface brightness. It is not our intention to reproduce the precise image of a CK.

%========================================================================
\section{Results}
%========================================================================

%========================================================================
\subsection{Diffuse Gas in the Helix Model} \label{ResHelixMod}
%========================================================================

\begin{figure}
\includegraphics[width=84mm]{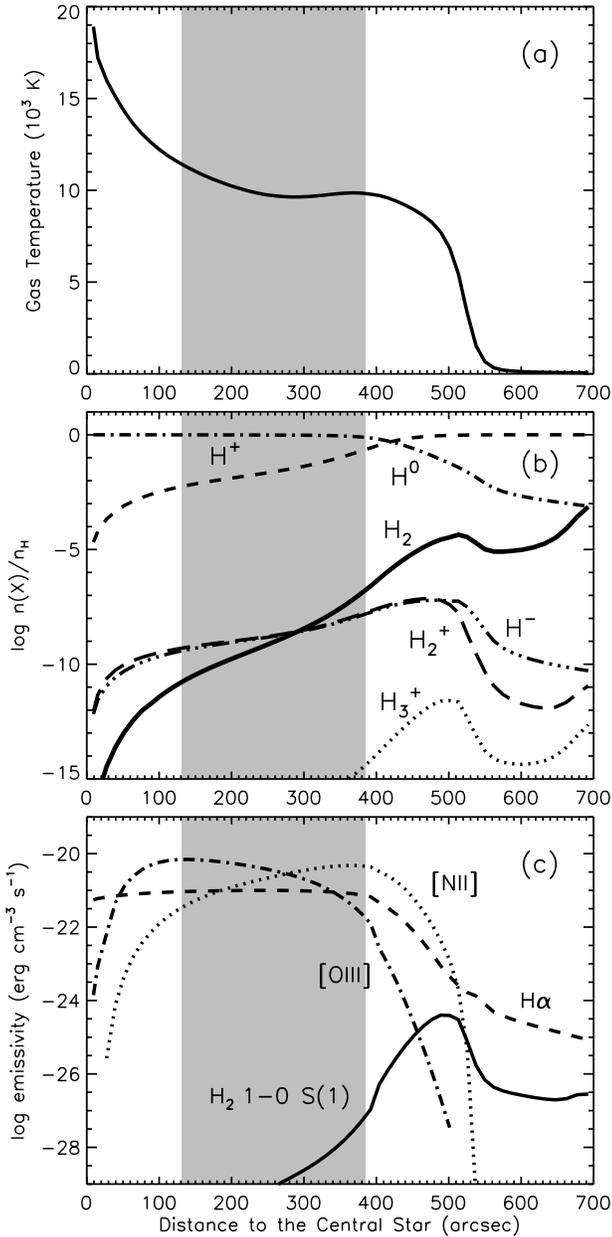}
\caption{Diffuse gas in the Helix model: (a) gas temperature, (b) density of H$^0$, H$^+$, H$^-$, H$_2$, H$_2^+$, and H$_3^+$; and (c) emissivity of some lines. CKs are not included in this model. The grey band represents the range of distances where the CKs are observed in the Helix; see discussion in the text.}
\label{HelixNoClumps}
\end{figure}

Figure \ref{HelixNoClumps} shows results for the Helix model obtained with the parameters of Table \ref{TabHelixPar}. The gas temperature profile is given in the top panel, while the radial profiles of H$^0$, H$^+$, H$^-$, H$_2^+$, and H$_3^+$ densities are plotted in the middle panel. Emissivities of some lines are shown in the bottom panel. The model in the figure applies to the diffuse gas only: it does not include CKs. The gas density and the dust-to-gas ratio are assumed to be uniform and equal to, respectively, $n_D = 50$~cm$^{-3}$ and $M_d/M_g = 10^{-3}$.

%------------------------------------------------------------------------
%------------------------------------------------------------------------
%------------------------------------------------------------------------
%------------------------------------------------------------------------

In the Helix, CKs are detected at distances from the central star (projected on the sky plane) between 132 arcsec and 384 arcsec \citep{Matsuura_etal_2009}. This interval is indicated by the gray band in Fig. \ref{HelixNoClumps}. The three-dimensional morphology must be taken into account to calculate the real distance. The morphology of the Helix was studied by \citet{Meaburn_etal_1998} and \citet{Odell_etal_2004}. According to \citet{Meaburn_etal_1998}, the helix structure (where the CKs are) has a toroidal shape, contained within an angle of about 20\degr on both sides of the equatorial plane of the nebula. The symmetry axis is inclined at about 37\degr with respect to the line of sight. \citet{Odell_etal_2004} found a slight different value to the inclination angle, 28\degr with respect to the line of sight. Taking this geometry into account, real distances must be greater than the projected distances. For example, a CK in a plane inclined by 37\degr from the line of sight could be up to 25 per cent farther from the central star than the projected distance. Even with this deprojection, the CKs are not much farther than 500 arcsec from the central star, that is, mostly inside the region where hydrogen is ionised. Our model also supports that the CKs coexist with the ionised species He$^{+}$, N$^{+}$,N$^{++}$, O$^{+}$, and O$^{++}$ in the diffuse gas. Our model has parameters similar to those of model 135/315 of \citet{Henry_etal_1999} and provides similar results for the ionisation structure.

According to \citet{Matsuura_etal_2009} and \citet{Meixner_etal_2005}, the H$_2$ emission of the Helix is associated with the CKs, although the instrumental resolution can not exclude a contribution from the diffuse gas in the outer ring. As can be seen in Fig. \ref{HelixNoClumps}, the calculated H$_2$ 1-0~S(1) emissivity from the diffuse gas is very low. Assuming a spherical gas distribution, the maximum H$_2$ 1-0~S(1) intensity is 10$^{-8}$~erg~cm$^{-2}$~s$^{-1}$~sr$^{-1}$, and the total 1-0~S(1) flux coming from inside the ionised region is 4$\times10^{-13}$~erg~cm$^{-2}$~s$^{-1}$. In contrast, the observed 1-0~S(1) flux of the Helix nebula is about 2$\times10^{-9}$~erg~cm$^{-2}$~s$^{-1}$ \citep[obtained from the observations published in][]{Speck_etal_2003}. The flux obtained with our model without GCs shows that the diffuse ionised gas contributes less than 0.02 per cent of the observed emission. Our calculation of the flux above does not take into account a possible increase in the average gas density in the rings or the shielding of the radiation by the CKs, which may increase the H$_2$ emission.

%========================================================================
\subsection{Physical and Chemical Properties of CKs}
%========================================================================

\begin{figure*}
\includegraphics[width=170mm]{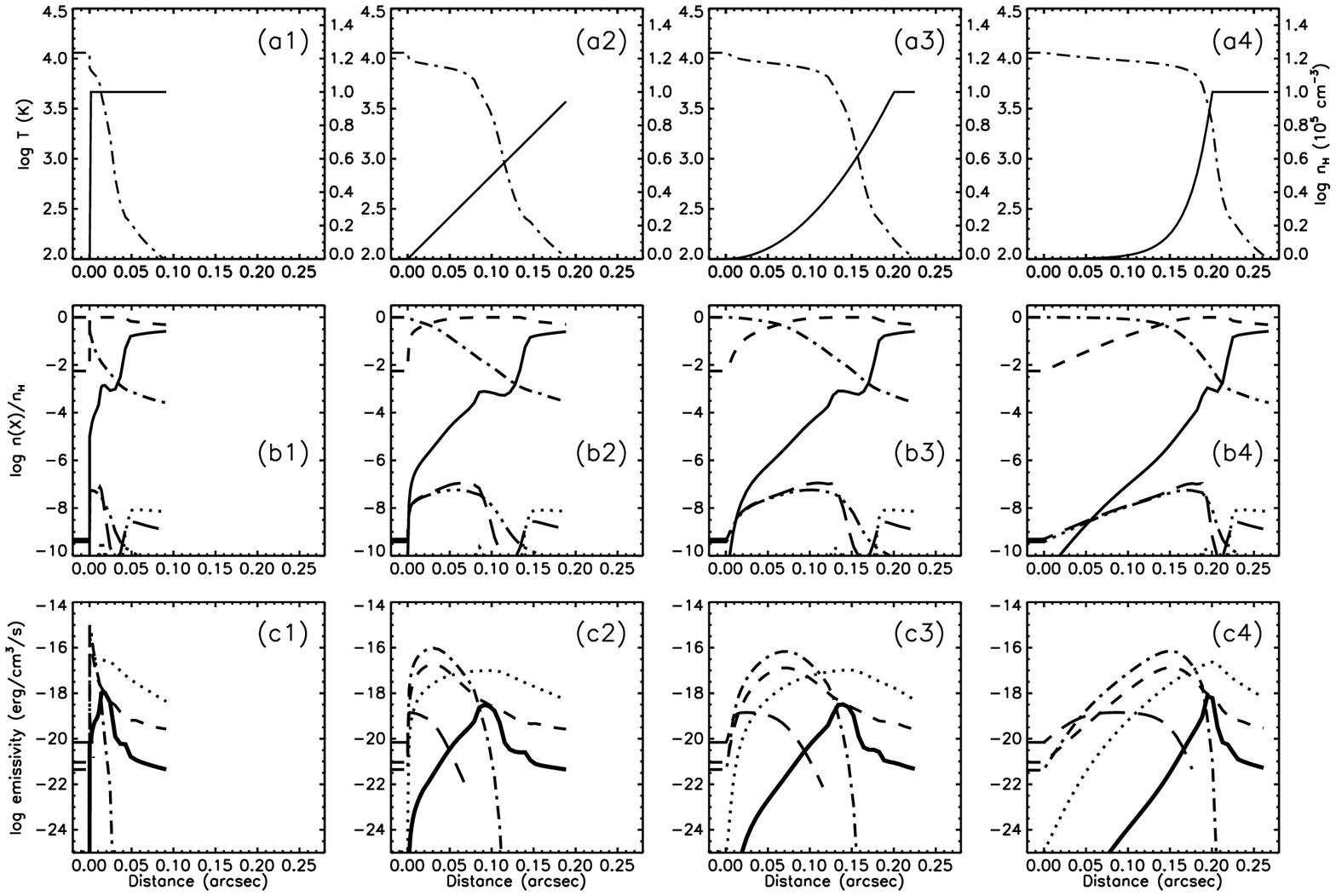}
\caption{Results for models K1 to K4 (left to right). Top: gas temperature (dot-dashed) and total density (solid) profiles. Middle: density of H$^0$ (dashed), H$^+$ (dot-dashed), H$^-$ (long dash), H$_2$ (solid), H$_2^+$ (three dot-dashed), and H$_3^+$ (dotted). Bottom: emissivities of the lines H$_2$ 1-0~S(1) (solid), H$\alpha$ (dashed), [\mbox{N\,{\sc ii}}] $\lambda$6583 (dot-dashed), [\mbox{O\,{\sc iii}}] $\lambda$5007 (long dashed), and [\mbox{O\,{\sc i}}] 63$\mu$m (dotted). Distance is taken from the border of the knot facing the star along the symmetry axis.}
\label{DensiK14}
\end{figure*}

\begin{figure*}
\includegraphics[width=170mm]{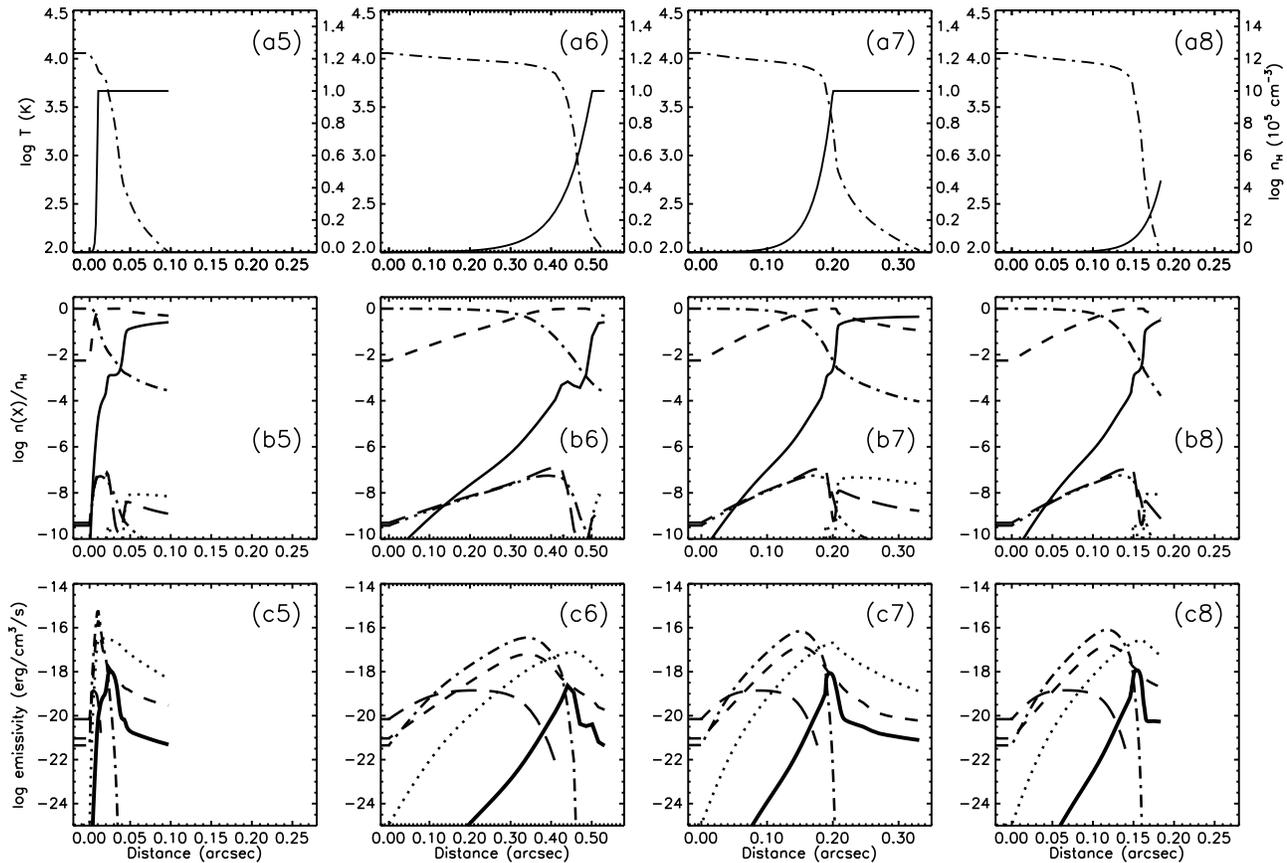}
\caption{Same as Fig. \ref{DensiK14} for models K5 to K8 (left to right). Note the different density scale for model K8 in panel (a8).}
\label{DensiK58}
\end{figure*}

\begin{figure*}
\includegraphics[width=170mm]{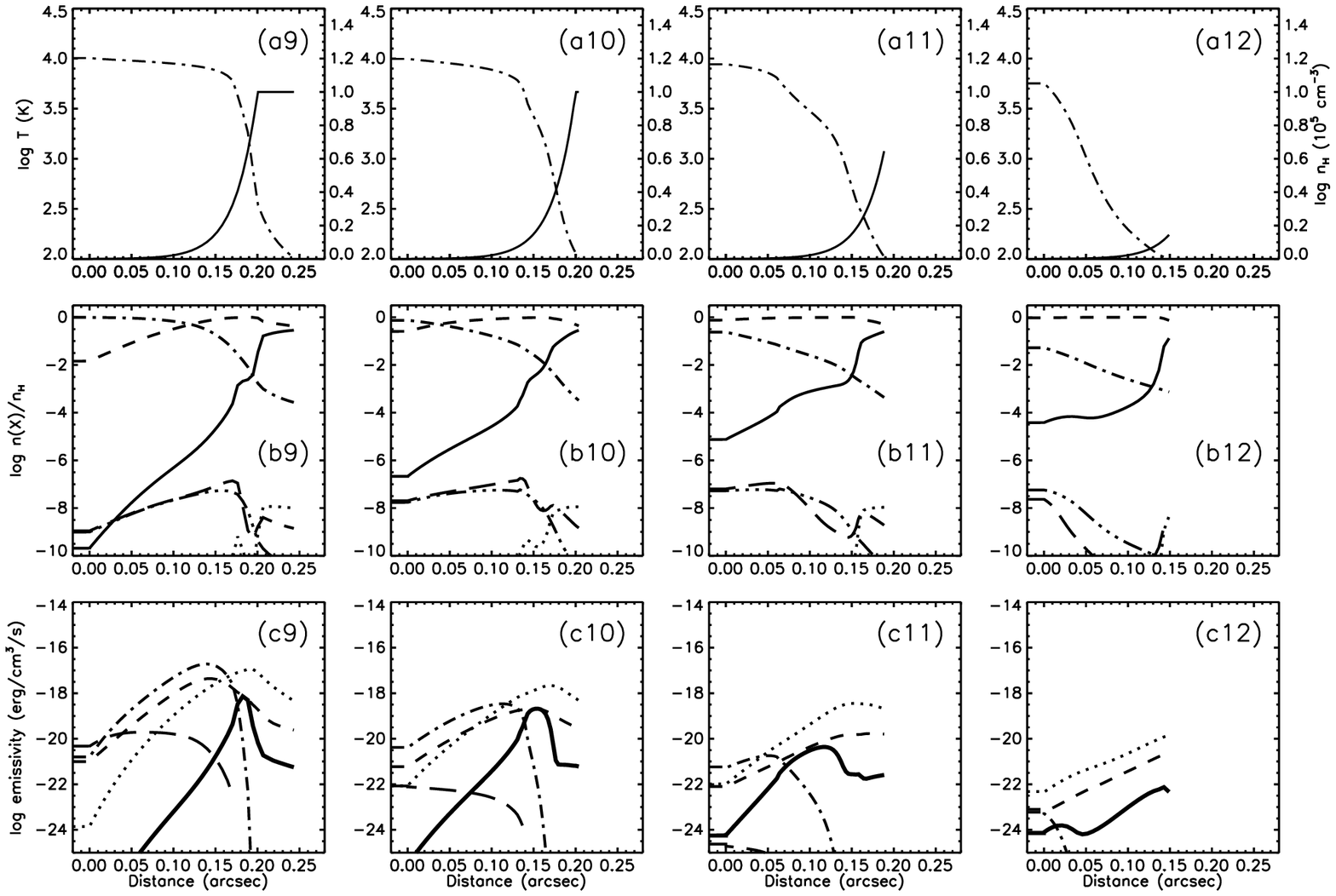}
\caption{Same as Fig. \ref{DensiK14} for models K9 to K12 (left to right).}
\label{DensiK912}
\end{figure*}

Figures \ref{DensiK14}, \ref{DensiK58}, and \ref{DensiK912} show results for models K1 to K12 (see the model parameters in Table \ref{TabKnotPar}). The upper panels show the gas temperature and the total H nuclei density profile; the middle panels show the density profiles of H$^0$, H$^+$, H$^-$, H$_2$, H$_2^+$, and H$_3^+$; and the bottom panels show the emissivity profiles of H$_2$ 1-0~S(1), H$\alpha$, [\mbox{N\,{\sc ii}}] $\lambda$6583, [\mbox{O\,{\sc iii}}] $\lambda$5007, and [\mbox{O\,{\sc i}}] 63$\mu$m.

%------------------------------------------------------------------------
%------------------------------------------------------------------------
%------------------------------------------------------------------------
%------------------------------------------------------------------------

%------------------------------------------------------------------------
%------------------------------------------------------------------------
%------------------------------------------------------------------------
%------------------------------------------------------------------------

%------------------------------------------------------------------------
%------------------------------------------------------------------------
%------------------------------------------------------------------------
%------------------------------------------------------------------------

\begin{figure}
\includegraphics[width=84mm]{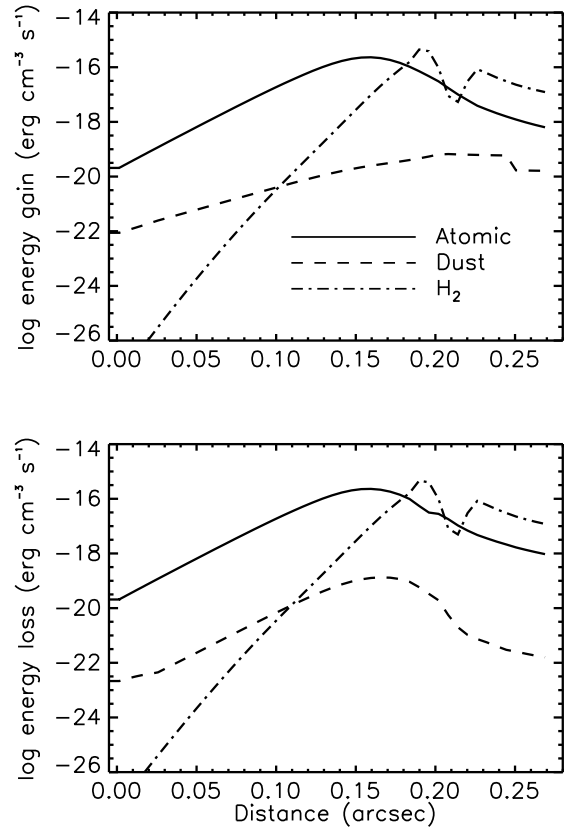}
\caption{Contribution of the mechanisms involving atoms, dust, and H$_2$ to the rates of energy gain and loss by the gas inside the CK model K4.}
\label{GainLoss}
\end{figure}

The gas temperature decreases with depth from the local diffuse gas value at the border of the CK to the 100~K limit assumed for our calculations. In the interface, the gas is hot, with temperatures $T >$ 2000~K. As previously mentioned, the balance between the energy gain and loss by the gas determine its temperature. Figure \ref{GainLoss} shows the rate of energy gain and loss inside the CK model K4. The results are analogue for other models. In the interface, atomic processes dominate the heating and cooling of the gas, as in the diffuse ionised gas around the CK. The H$_2$ molecule dominates the heating and cooling in the core of the CK, while atomic mechanisms contributes to about 10 per cent of the total rates of energy loss and gain. Collisional excitation and de-excitation of H$_2$ are the most important mechanisms of loss and gain, respectively. Photoionisation of H$_2$ may also contribute to the energy gain ($<$10 per cent). Heating by the photoelectric effect on grains may be significant only if $M_d/M_g >> 10^{-2}$. For example, this mechanism contributes less than one per cent to the total energy input in model K4 ($M_d/M_g = 10^{-3}$), and less than 5 per cent for model K7 ($M_d/M_g = 10^{-2}$). The calculated grain temperature inside the CKs is approximately 20~K.

%------------------------------------------------------------------------
%------------------------------------------------------------------------
%------------------------------------------------------------------------
%------------------------------------------------------------------------

The transition between ionised and neutral H occurs in the interface, around $T \sim$ 8000~K. Below this temperature and down to 100~K, hydrogen is mostly neutral. The density of molecular hydrogen increases inward within the CK, reaching a fractional abundance in excess of 30 per cent. In cases with a high dust-to-gas ratio (see Sect. \ref {effectpar}), a fully molecular region is produced while the temperature is still above 100~K. For other dust-to-gas ratios, the fully molecular region must have lower temperatures. The relative densities of H$^-$, H$_2^+$, and H$_3^+$ are smaller than $10^{-7}$ in all the models.

\begin{figure}
\includegraphics[width=84mm]{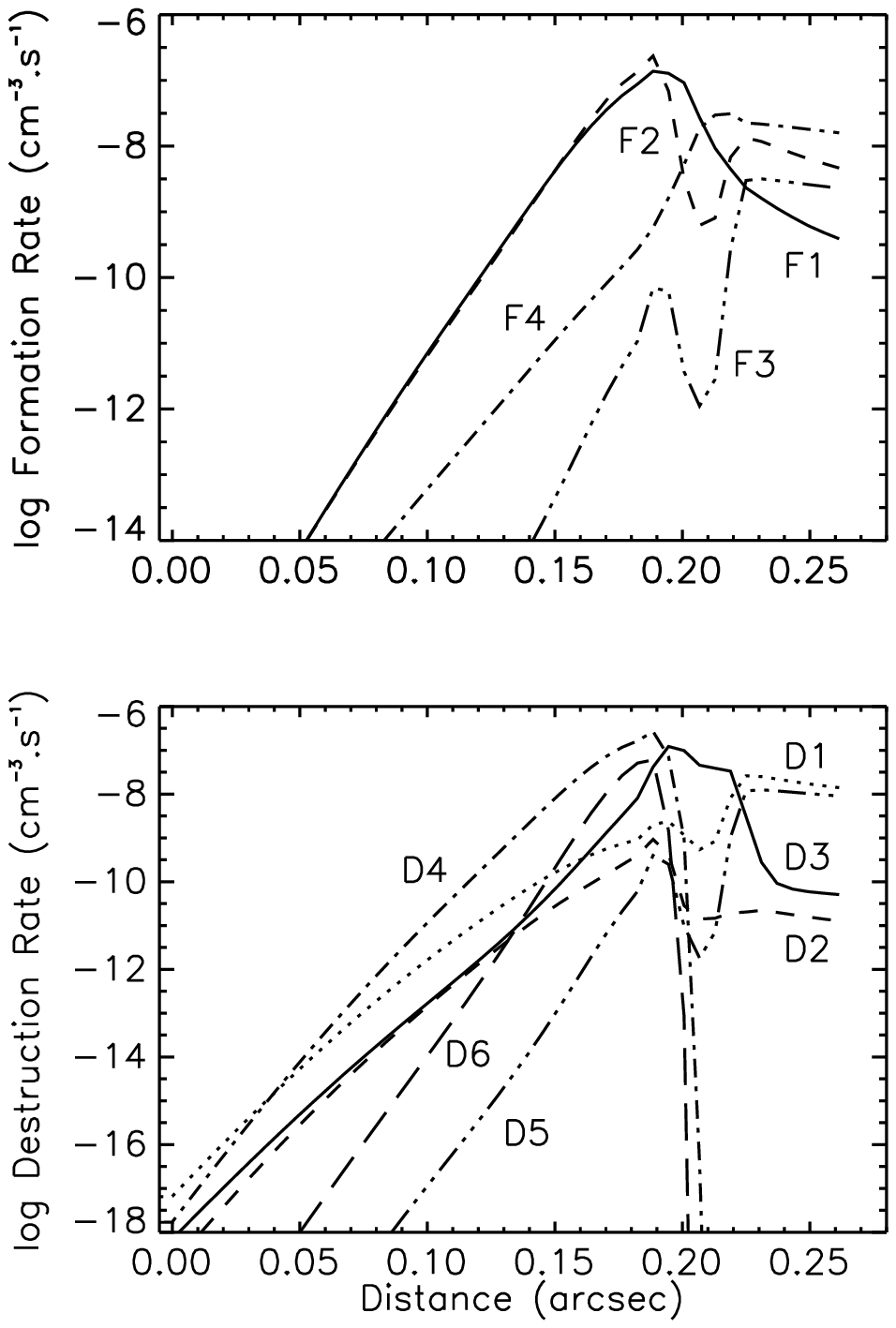}
\caption{Rates of the important processes of H$_2$ formation and destruction inside the CK model K4. See text for the reaction codes.}
\label{taxasEQ}
\end{figure}

According to our models, inside a CK the most important H$_2$ formation mechanisms are:

\begin{description}
\item (F1) H + H$^-$ $\rightarrow$ H$_2$ + e$^-$		% (15)
\item (F2) H$_2^+$ + H $\rightarrow$ H$^+$ + H$_2$	% (16)
\item (F3) H$_3^+$ + e$^-$ $\rightarrow$ H$_2$ + H	% (13)
\item (F4) 2H + grain $\rightarrow$ H$_2$ + grain 	% (25)
\end{description}
and the main H$_2$ destruction mechanisms are:

\begin{description}
\item (D1) H$_2$ + h$\nu$ $\rightarrow$ H$_2^+$ + e$^-$		% (2)
\item (D2) H$_2$ + h$\nu$ $\rightarrow$ 2H					% (3)
\item (D3) H$_2$ + h$\nu$ $\rightarrow$ H$_2^*$ $\rightarrow$ 2H	% (4)
\item (D4) H$_2$ + H$^+$ $\rightarrow$ H + H$_2^+$			% (17)
\item (D5) H$_2$ + H$_2^+$ $\rightarrow$ H$_3^+$ + H			% (24)
\item (D6) H$_2$ + H $\rightarrow$ 3H						% (27)
\end{description}
Both neutral and ionised species are involved in the H$_2$ formation and destruction processes; UV photons are important for the destruction of the molecule. Figure \ref{taxasEQ} shows the rate of these processes as a function of the distance inside a CK. The associative detachment (F1) and the charge exchange (F2) reactions dominate the formation of H$_2$ at the border of the CK facing the star. The associative detachment is less important towards high depths, while the formation on grain surfaces becomes more important and dominates the H$_2$ formation in the CK core. The charge exchange and ion-molecule (F3) reactions contribute significantly to the molecular formation in the whole studied region.

%------------------------------------------------------------------------
%------------------------------------------------------------------------
%------------------------------------------------------------------------
%------------------------------------------------------------------------

The list of important reactions of H$_2$ formation and destruction in CKs is similar to the important reactions for the diffuse ionised gas \citep[see][]{Aleman_Gruenwald_2004}. The differences are: (F3) and (D5) are not important in the diffuse gas, and the dissociation of H$_2$ by electron collision is not important inside the CKs.

Models by \citet{Henney_etal_2007} provide almost the same list of significant reactions for H$_2$, with the exception of destruction of H$_2$ by the reaction H$_2$ + e$^- \rightarrow$ H$^-$ + H, instead of (D6), probably because in their models the advection brings the ionisation front closer to the molecular zone, increasing the density of electrons and decreasing the density of H$^0$ coexisting with H$_2$.

Estimates of the timescale for H$_2$ formation/destruction can be obtained with the formula

\begin{equation}
   \Delta t = \frac{n(H_2)}{dn(H_2)/dt},
\end{equation}
where $\Delta t$ is the characteristic timescale, $n($H$_2)$ is the molecular hydrogen density, and d$n$/d$t$ is the formation/destruction rate. Our equilibrium models give timescales inside a CK typically from one year at the border to $10^4$ years in the core of a CK.

%========================================================================
\subsubsection{Effect of the Input Parameters} \label{effectpar}
%========================================================================

\begin{figure}
\includegraphics[width=84mm]{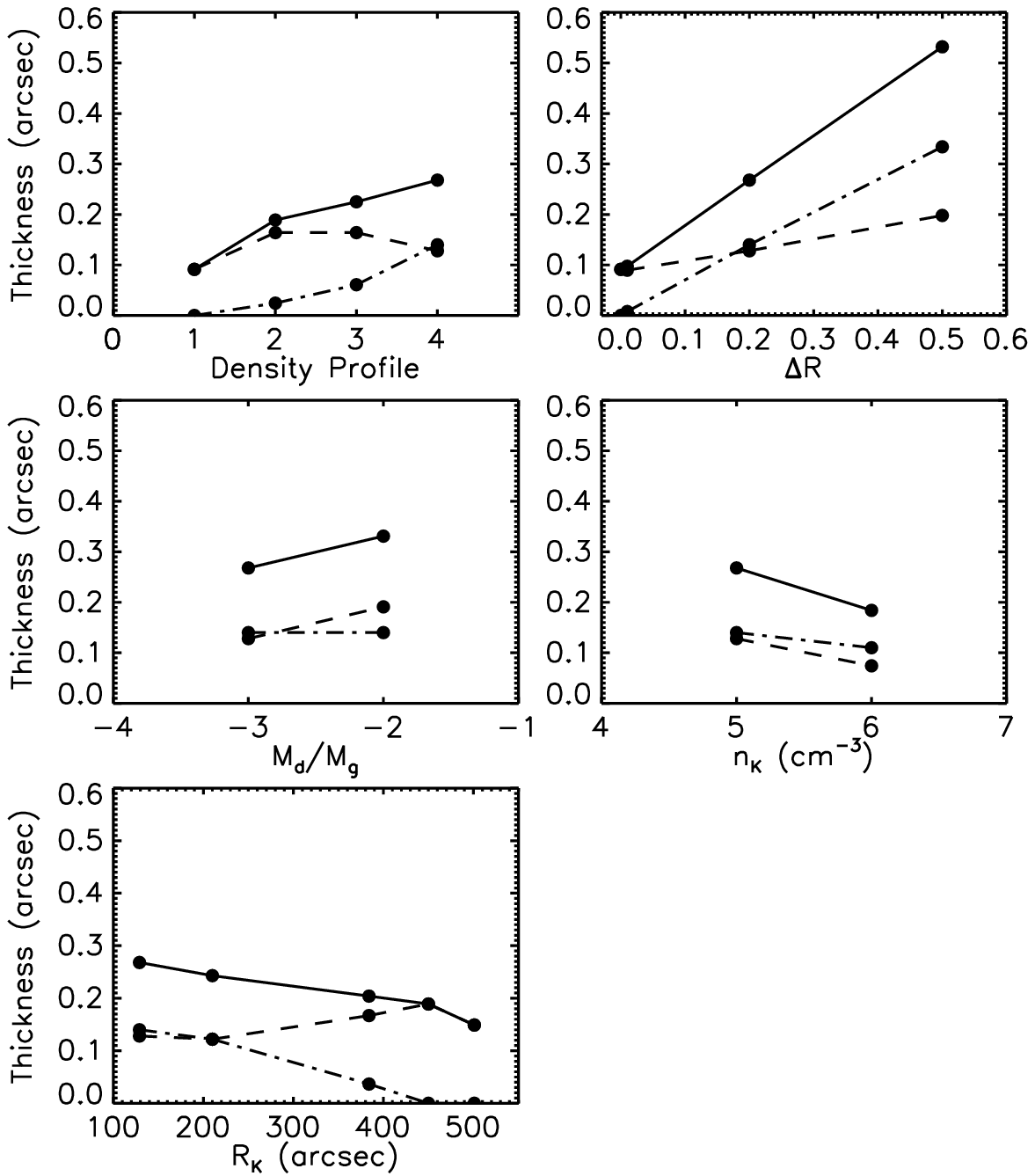}
\caption{Effect of the various input parameters on the H ionisation structure. 
In each plot, the total thickness from the border of the knot to the position where $T =$ 100~K ($\Delta R_T$) is represented by the solid line. The widths of the ionised and neutral zones are represented by the dot-dashed and the dashed curves, respectively.}
\label{DepPar}
\end{figure}

As shown in Figs. \ref{DensiK14} to \ref{DensiK912}, models with different input parameters yield different ionisation structures and hence result in different line emissivity profiles. Figure \ref{DensiK14} shows models with different CK density profiles (models K1 to K4). Using the parameters of the model K4 as reference, in model K5 and K6 we change the interface thickness; in model K7, the dust-to-gas ratio; in model K8, the core density (Fig. \ref{DensiK58}).  Models K9 to K12 have different distances from the central star (Fig. \ref{DensiK912}). Figure \ref{DepPar} summarises the effect of the input parameters on the total thickness from the border of the knot to the position where $T =$~100~K (i.e., $\Delta R_T$). The figure also shows the thickness of ionised and neutral zones of this region. Here the boundary between the ionised and neutral zones is assumed to be where H ionization degree is 0.5.

%------------------------------------------------------------------------
%------------------------------------------------------------------------
%------------------------------------------------------------------------
%------------------------------------------------------------------------

The thickness of the CK ionised zone increases significantly from type~1 to type~4 interface. The variation of the thickness of the neutral zone is less significant, but it also changes with the type of the interface. The model with type~1 interface has the smallest absolute thickness, while type~2 has the greatest value. As a result, $\Delta R_T$ increases significantly from type~1 to type~4 interface.

The density profile in the interface may result from the physical processes occurring and the nature of knots. For example, an $r^2$ function would be expected in the case of a spherical equilibrium outflow \citep{Burkert_Odell_1998}. According to \citet{Burkert_Odell_1998} the H$\alpha$ surface brightness profile is better fitted by an exponential function than by an $r^3$ function (resulting from an $r^2$ increase in density). Following the results of \citet{Burkert_Odell_1998}, we assume an exponential increase hereafter.

The thickness of the ionised zone and $\Delta R_T$ increase for models with thicker interfaces (greater $\Delta R$), as can be seen by comparing models K5, K4, and K6 (Figs. \ref{DensiK14} and \ref{DensiK58}). The three models have the same parameters, except for $\Delta R$, which takes, respectively, the values 0.01, 0.2, and 0.5. In models with $\Delta R=0.01$ the neutral region is about 10 times larger than the ionised, while for $\Delta R=0.2$ they are about the same size. 

The rate of H$_2$ formation on grain surface is proportional to the dust-to-gas ratio. The contribution of dust to the gas energy input also increases in models with higher $M_d/M_g$, producing a more extended region with 100 $< T <$ 300~K. This allows hydrogen to become fully molecular in such a region. Compare models K4 (Fig. \ref{DensiK14}) and K7 (Fig. \ref{DensiK58}), which differ only by the value of dust-to-gas ratio (the dust-to-gas ratio for K7 is 10 times larger than for K4).

Density and gas temperature profiles for model K8 is shown in Fig. \ref{DensiK58}. This model is similar to model K4, but with a higher core density $n_K$. Comparing both models, it can be seen that both ionised and neutral regions reduce in size in the denser model. Several mechanisms can contribute to this difference, such as the increase in the radiation extinction by gas and dust, increase in the H$_2$ formation on grain surface, heating of the gas. It is important to notice that, since we keep the dust-to-gas ratio constant throughout the model, an increase in the gas density implies in an increase of the dust density.

An important parameter to the ionisation structure of the CKs is the distance from the central star, since the ionising flux and spectrum may change significantly with the position in the nebula. CKs farther from the central star have smaller ionised zones. If the CK is beyond the Helix ionisation front there is no ionised region, and the knot is completely neutral (see model K12 in Fig. \ref{DensiK912}).

%========================================================================
\subsection{Warm H$_2$ 1-0~S(1) Emission}
%========================================================================

As can be seen in Figs. \ref{DensiK14} to \ref{DensiK912} the emissivity of the H$_2$ 1-0~S(1) line in the CKs is high in a warm region, where the temperature ranges between 300 and 7000~K. The peak in the 1-0~S(1) emissivity in the studied region occurs where the density is around 40 per cent of the core density (with the exception of interface type 1 models, which have no intermediate density). This is also true for other rovibrational lines. This component of the H$_2$ emission can explain the excitation temperatures around 900-1800~K found by \citet{Cox_etal_1998} and \citet{Matsuura_etal_2007}. The spatial width of the peak in the H$_2$ 1-0~S(1) emissivity is similar for all models, but the peak emissivity decreases with the distance from the central star.

Inside the CKs, the H$_2$ rovibrational levels of the ground electronic state are mainly excited and de-excited by collisional processes, particularly for lower $J$ levels. Formation pumping may contribute to high $J$ levels population ($J >$ 10), in the region where H is neutral. In such a region, rovibrational radiative de-excitation is an important mechanism for the population of levels with $J >$ 15 or for lower $J$ levels with $v \sim$ 5. UV Pumping is important for the population of levels with $v >$ 4 and lower $J$.

\begin{figure}
\includegraphics[width=84mm]{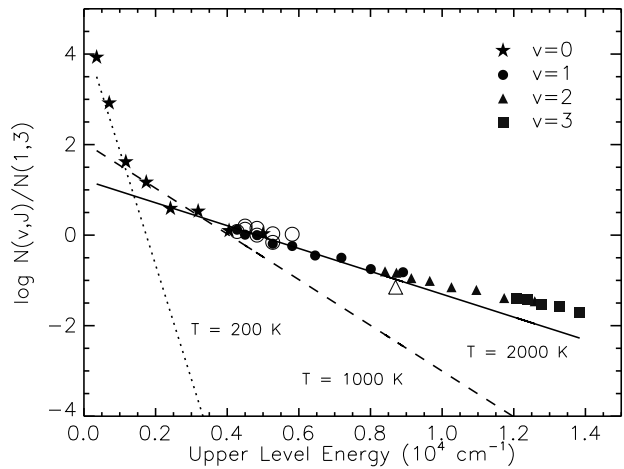}
\caption{H$_2$ excitation diagram. The effective column density was calculated from the surface brightness at the peak intensity of 1-0~S(1). Open symbols represent observations and filled symbols represent models. Lines are Boltzmann distributions for the temperature indicated near each curve.}
\label{ExcitH2}
\end{figure}

Figure \ref{ExcitH2} shows the H$_2$ excitation diagram for model K9. The effective column densities of the rovibrational levels $v =$ 0, 1, 2 and 3 were calculated using the intensities determined in our model: these are indicated by the filled symbols. The lines in Fig. \ref{ExcitH2} represent Boltzmann distributions for three different temperatures. Different regions within the CK contribute to different parts of this plot. The highest energy levels are populated by collisions in the hottest regions of the CK. The lower energy levels are populated throughout the knot, at a range of temperatures. As a result, the integrated spectrum yields a much higher temperature for transitions involving a high upper energy level. The plot clearly shows these different excitation components of the H$_2$ emission. The lines of the band 1-0 and 2-1 are well represented by an excitation temperature of approximately 2000~K. The 0-0 band shows a distribution of lower temperatures, approximated in the plot as two components, with excitation temperatures of 900 and 200 K.

%------------------------------------------------------------------------
%------------------------------------------------------------------------
%------------------------------------------------------------------------
%------------------------------------------------------------------------

Effective column densities obtained from peak intensities measured by \citet{Matsuura_etal_2007} (their region a) are shown in Fig. \ref{ExcitH2} (open symbols). The agreement between the excitation temperatures of the model and the observations is evident. \citet{Matsuura_etal_2007} found a temperature of 1800~K from the 1-0 and 2-1 bands in a knot relatively close to the star, well represented by the model for these bands. (Note that the observed knot is 150 arcsec from the star, while the model is for a (deprojected) distance of 210 arcsec.)

For the 0-0 band, \citet{Cox_etal_1998} found from ISO observations that the observed column densities obtained from lines S(2) to S(7) are well represented by Boltzmann distribution at a temperature around 900 K. This agrees with the excitation temperature of 1000~K predicted by the current model. 

\citet{Matsuura_etal_2007} suggested that the different excitation temperatures they found (inferred from rovibrational lines) and the value obtained by \citet{Cox_etal_1998} (from pure rotational lines) is evidence for temperature variation within the nebula. \citet{Hora_etal_2006} found some indication for a gradient in the excitation in the Helix CKs with the distance from the central star, but in our models, a significant gradient is expected only at large distances, close to the ionisation boundary of the PN. Instead, the different excitation temperatures found by \citet{Cox_etal_1998} and \citet{Matsuura_etal_2007} can be explained by the effect of the excitation within a single CK.

Using the ratios between lines of the S and Q branches of band 1-0 lines departing from the same level and assuming an extinction law of the form $\lambda^{-1.7}$ \citep[see][]{Davis_etal_2003}, we estimate an extinction of about 10-20 per cent at 2$\mu$m. The model assumes that the knot is fully transparent, while in reality a fraction of the emission from the backside of the knot is extincted. However, this will not have a major effect on our results.

For comparison with H$_2$, the emissivity of a few intense atomic lines are also shown in Figs. \ref{DensiK14} to \ref{DensiK912}. The [\mbox{N\,{\sc ii}}] $\lambda$6583 emission is produced in the interface of the CKs. H$\alpha$ is also produced in the interface region, but there is a less intense extended emission region produced by the high energy photons (which ionise the H atoms; the following recombination produces the line). The emissivity of the [\mbox{N\,{\sc ii}}] line loses importance for models farther from the central star. For such models, the [\mbox{N\,{\sc ii}}] line is produced only near the border of the CK and its emissivity profile becomes very different from that of H$\alpha$. Such behaviour may explain the separation between the H$\alpha$ and [\mbox{N\,{\sc ii}}] peaks observed by \citet{Odell_etal_2000}.The decrease with distance of the [\mbox{N\,{\sc ii}}] emissivity is stronger than for the 1-0~S(1) line, which can explain why all CKs with [\mbox{N\,{\sc ii}}] emission have a 1-0~S(1) counterpart, whilst the opposite is not true, as observed by \citet{Matsuura_etal_2009}.

Inside the CKs, the [\mbox{O\,{\sc iii}}] $\lambda$5007 emission is only produced in the interface and in models closer to the central star. In such cases, the emissivity inside the CK increases only up to one order of magnitude from the diffuse gas value, a small value when compared to other lines, such as H$\alpha$ or [\mbox{N\,{\sc ii}}] $\lambda$6583. Images of the Helix show [\mbox{O\,{\sc iii}}] emission in the diffuse gas and, in some cases, in the tip of CKs \citep{Odell_etal_2007}. According to \citet{Walsh_Meaburn_1993}, 40 per cent of the CKs does not show any emission in [\mbox{O\,{\sc iii}}] images.

The predicted emissivity of the IR fine structure line [\mbox{O\,{\sc i}}] 63$\mu$m inside the CKs is very high as can be seen in Figs. \ref{DensiK14}, \ref{DensiK58}, and \ref{DensiK912}. Reay et al (1988) found a correlation between the H$_2$ 1-0~S(1) and the optical [OI] $\lambda$6300 in a sample with several PNe, including the Helix. The optical line is produced only in the interface, while the profile of the IR [OI] line is more similar to the one of 1-0~S(1).

As discussed before, if the CK is beyond the Helix ionisation front there is no ionised region, and the knot is completely neutral (see models K11 and K12 in Fig. \ref{DensiK912}). In this case, the emissivity of both [\mbox{N\,{\sc ii}}] and [\mbox{O\,{\sc i}}] lines mentioned above is very low in the CKs and would not be detected.

%========================================================================
\subsection{H$_2$ 1-0~S(1) Surface Brightness} %========================================================================

\begin{figure}
\includegraphics[width=84mm]{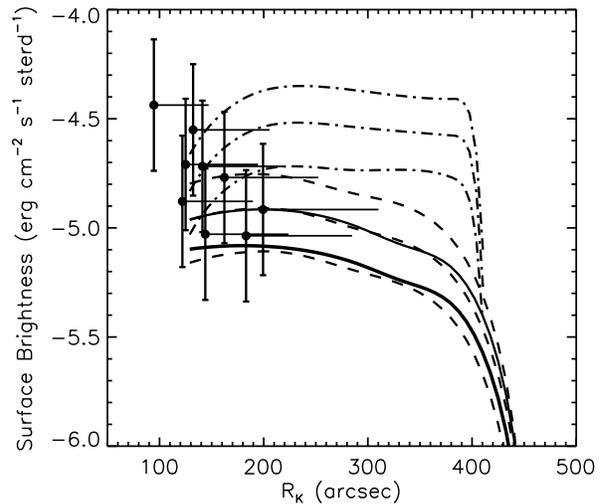}
\caption{H$_2$ 1-0~S(1) surface brightness of a cometary knot as a function of the distance to the central star. Sets of solid, dashed, and dot-dashed curves represent models with $\Delta R =$ 0.5, 0.2, and 0.01 arcsec, respectively. Different curves within each set represents CK radius of 0.5, 1.0, and 2.0 arcsec, with the surface brightness increasing for larger CK radius. Dots represent measured values. The uncertainty in the distance is estimated assuming that the Helix symmetry axis is inclined 37$\,^{\circ}$ with respect to the line of sight}
\label{SurfBrightH2}
\end{figure}

Measurements of the H$_2$ 1-0~S(1) line surface brightness of some representative CKs in the Helix nebula are plotted in Fig. \ref{SurfBrightH2}. We identified 10 isolated CKs detected both in H$\alpha$ and H$_2$ images. We measured 2.12\,$\mu$m H$_2$ intensities from the images obtained by \citet{Matsuura_etal_2009}. To calibrate the intensities, we used five stars within the observed field to measure the zero-point. We assume that the 2MASS $K'$-magnitude of these stars are the same as the magnitudes in H$_2$ filter. We apply a 25-pixel radius for aperture photometry and take the 35--50 pixel ring for the background measurements. The pixel scale is 0.117 arcsec. Table \ref{SBTable} lists the knots positions and the measurements.

Curves in Figure \ref{SurfBrightH2} show the calculated H$_2$ 1-0~S(1) surface brightness of CK models as a function of the CK distance to the central star. The calculated surface brightness was averaged over the same aperture as the measurements to allow direct comparison. Different curves represent different $\Delta R$ and CK radius. The surface brightness increases with a decrease of $\Delta R$, an increase of the CK radius. CKs closer to the central star tend to have higher 1-0~S(1) surface brightness. The decrease in the H$_2$ 1-0~S(1) surface brightness with distance from the central star was also noticed by \citet{Odell_etal_2007}. If the CK is beyond the Helix ionisation front, the 1-0~S(1) surface brightness drops dramatically, since there would be no enough radiation or temperature to excited significantly the upper vibrational levels of the molecule. Models can reproduce the magnitude of observed surface brightness and its decrease with distance to the central star.

%------------------------------------------------------------------------
%------------------------------------------------------------------------
%------------------------------------------------------------------------
%------------------------------------------------------------------------

%------------------------------------------------------------------------
\begin{table}
\centering
\begin{minipage}{140mm}
\caption{H$_2$ 1-0~S(1) Surface Brightness Measurements.}
\begin{tabular}{@{}ccccc@{}}
\hline
RA (E2000) & $\delta$ (E2000) & Distance & Peak Surface Brightness \\
(h m s) & (\degr \arcmin \arcsec) & (arcsec) & 
(erg cm$^{-2}$ s$^{-1}$ sterd$^{-1}$) \\
\hline
22:29:45.987 & -20:49:05.98 & 125 & 2.0 $\times 10^{-5}$ \\
22:29:42.852 & -20:49:01.39 & 95  & 3.7 $\times 10^{-5}$ \\
22:29:42.873 & -20:47:43.35 & 162 & 1.7 $\times 10^{-5}$ \\
22:29:47.370 & -20:47:37.92 & 199 & 1.2 $\times 10^{-5}$ \\
22:29:40.616 & -20:47:52.70 & 144 & 9.6 $\times 10^{-6}$ \\
22:29:41.285 & -20:47:14.67 & 183 & 9.2 $\times 10^{-6}$ \\
22:29:37.829 & -20:48:01.66 & 132 & 2.8 $\times 10^{-5}$ \\
22:29:42.820 & -20:48:27.77 & 122 & 1.3 $\times 10^{-5}$ \\
22:29:39.823 & -20:47:53.55 & 141 & 1.9 $\times 10^{-5}$ \\
\hline
\label{SBTable}
\end{tabular}
\end{minipage}
\end{table}
%------------------------------------------------------------------------

The 1-0~S(1) peak brightness is slightly higher for models with higher dust-to-gas ratio. The increase caused by changing the dust-to-gas ratio from 10$^{-3}$ to 10$^{-2}$ is about 20 per cent. The 1-0~S(1) peak brightness also increases with increasing $n_K$. The difference between models with $n_K =$ 10$^{5}$ and 10$^{6}$ cm$^{-3}$ is up to 40 per cent (20 per cent in the models farther from the central star).

\citet{Matsuura_etal_2009} show that, for the same number of CKs, the outer ring shows twice the surface brightness as the inner ring. According to them, such difference may be caused by the diffuse gas or by the difficulty in isolating CKs in the images. Since we do not expect a significant contribution from the diffuse gas and the surface brightness decreases with the distance to central star, the second explanation seems more likely.

%========================================================================
\section{Conclusion}
%========================================================================

As discussed in the Introduction some features of the CKs emission indicate that they are significantly affected by the radiation field produced by the Helix central star. In this paper we use the photoionisation code \textsc{Aangaba} to assess the contribution that photoionisation may have on the molecular hydrogen emission, supporting that photoionisation plays a major role in such objects, which is in agreement with previous results from \citet{Odell_etal_2007} and \citet{Henney_etal_2007}.

Our Helix nebula model with no CKs indicates that emission of H$_2$ from ionised diffuse gas is not significant, agreeing with the observations that the H$_2$ emission of the Helix is associated with the CKs \citep{Matsuura_etal_2009, Meixner_etal_2005}

We study CK models with different sets of input parameters, chosen to cover the range of values inferred from Helix observations. The important mechanisms of gas heating and cooling, formation and destruction of H$_2$, excitation and de-excitation of H$_2$ were determined.

The emissivity of the H$_2$ 1-0~S(1) line in the CKs is important in a region with temperatures between 300 and 7000~K. Such a warm component of the H$_2$ emission may explain the excitation temperatures of around 900-1800~K found by \citet{Cox_etal_1998} and \citet{Matsuura_etal_2007}. The contribution of this warm region to the emission of rovibrational lines is very important. For pure rotational lines of the $v =$~0 level the contribution of the colder regions should be significant. The excitation diagrams obtained from our models seems to agree very well with the observations. Comparison with measurements indicates that models can reproduce the magnitude of observed surface brightness.

An important parameter to the ionisation structure of the CKs is the distance from the central star, since the ionising spectrum may change significantly with the position in the nebula. CKs farther from the central star have smaller ionised zones. If the CK is beyond the Helix ionisation front there is no ionised region, the knot is completely neutral and there would be no enough radiation or temperature to excited significantly the upper vibrational levels of the molecule. The 1-0~S(1) intensity would be very low.

Our results also indicates that the separation between the H$\alpha$ and [\mbox{N\,{\sc ii}}] peaks observed by \citet{Odell_etal_2000} may be an effect of the distance of the knot from the star, since for knots farther from the central star the [\mbox{N\,{\sc ii}}] line is produced closer to the border of the CK than H$\alpha$.

The decrease with distance of the [\mbox{N\,{\sc ii}}] emissivity is stronger than for 1-0~S(1) line, which can explain why all CKs with [\mbox{N\,{\sc ii}}] emission have a 1-0~S(1) counterpart, but the opposite is not true, as observed by \citet{Matsuura_etal_2009}.

As pointed out by \citet{Burkert_Odell_1998}, the interface between the diffuse gas and the CK core may provide important clues about the mechanisms that shape and sustain the PNe CKs. Our models show that there are significant differences in the models results depending on the assumed density profiles of this region. Images that could resolve this region in great detail are then essential.

We find that the temperature derived from H$_2$ observations even of a single knot, will depend very strongly on the observed transitions, with much higher temperatures derived from excited levels. This is caused by the large range of temperatures present, and the presence of significant amounts of molecular hydrogen within the mini-PDR. This explains the puzzling temperature differences found by previous observers. 

We also find that H$_2$ required to fit the observations, is consistent with the abundances calculated through equilibrium chemistry, with a variety of formation and dissociation reactions. There is no need to presume that the molecular hydrogen in the knots predate the ionisation of the nebula, assuming that there is enough time to reach equilibrium.  This does not resolve the issue of when and how the knots formed, but we cannot exclude models where the knots formed after the onset of ionisation, nor models where the knots formed already in the AGB wind.

%========================================================================
\section*{Acknowledgments}
%========================================================================

I.A. acknowledged the financial support from CNPq Brazil PDE fellowship number 201950/2008-1 and STFC rolling grant ST/F003196. I.A. is grateful to the staff and students of the Astrophysics Group of University College of London for the hospitality during her visit in May 2010. We thank Paul Woods for discussions about the chemistry of C and O rich planetary nebulae, and the anonymous referee for the valuable suggestions to improve this paper.

%========================================================================

%========================================================================
\bibliographystyle{mn2e}  % style mn2e.bst
\bibliography{knots}      
%========================================================================

\bsp

\label{lastpage}

\end{document}